\documentclass[prb,aps,epsf,twocolumn,showpacs,10pt]{revtex4-1}
\usepackage{graphicx,amsfonts,times,bm,amsmath,verbatim,color,array}

\renewcommand{\)}{\right)}

\begin{document}
\title{Chiral anomaly and longitudinal magnetotransport in type-II Weyl semimetals}

\author{Girish Sharma$^{1,2}$}
\author{Pallab Goswami$^3$}
\author{Sumanta Tewari$^{1,4}$}

\affiliation{$^1$Department of Physics and Astronomy, Clemson University, Clemson, SC 29634, USA\\
$^2$Department of Physics, Virginia Tech, Blacksburg, VA 24061, U.S.A\\
$^3$Condensed Matter Theory Center and Joint Quantum Institute, Department of Physics, University of Maryland, College Park, Maryland 20742- 4111 USA\\
$^4$Department of Physics, Indian Institute of Technology, Kharagpur, Kharagpur, 721302, India}

\begin{abstract}
In the presence of parallel electric and magnetic fields, the violation of separate number conservation laws for the three dimensional left and right handed Weyl fermions is known as the chiral anomaly. The recent discovery of Weyl and Dirac semimetals has paved the way for experimentally testing the effects of chiral anomaly via magneto-transport measurements, since chiral anomaly can lead  to negative longitudinal magneto-resistance (LMR) while the transverse magneto-resistance remains positive. More recently, a type-II Weyl semimetal (WSM) phase has been proposed, where the nodal points possess a finite density of states due to the touching between electron- and hole- pockets. It has been suggested that the main difference between the two types of WSMs (type-I and type-II) is that in the latter, chiral anomaly induced negative LMR (positive longitudinal magnetoconductance) is strongly anisotropic, vanishing when the applied magnetic field is perpendicular to the direction of tilt of Weyl fermion cones in a type-II WSM. We analyze chiral anomaly in a type-II WSM in quasiclassical Boltzmann framework, and find that the chiral anomaly induced positive longitudinal magneto-conductivity is present  along any arbitrary direction. Thus, our results are pertinent for uncovering transport signatures of type II WSMs in different candidate materials.
\end{abstract}

\maketitle


\section{Introduction} The celebrated massless Dirac and Weyl equations were originally introduced for describing fundamental particles in high energy physics~\cite{Peskin}. However, in recent years they have transcended the barrier of high energy physics and become relevant for describing emergent, linearly dispersing, low energy excitations of several condensed matter systems~\cite{Murakami1:2007, Murakami2:2007, Yang:2011, Burkov1:2011, Burkov:2011, Volovik, Wan:2011, Xu:2011}. The Weyl equation captures the touching of two nondegenerate bands at isolated points in the momentum space, and these diabolic points act as the source and sink of Abelian Berry curvature. Consequently, Weyl semimetals violate spatial inversion (SI) or time reversal (TR) symmetry~\cite{Wan:2011, Xu:2011, Burkov:2011, Volovik}. 
The low energy effective Hamiltonian around a Weyl point $\mathbf{K}$ in the momentum space can be written as
\begin{eqnarray}
H_{\mathbf{k}} = \hbar \sum_{j=1}^{3} \; v_j (\mathbf{k}_j- \mathbf{K}_j)\sigma_j,
\label{Eq_H_k_weyl_1}
\end{eqnarray}
where $\sigma_j$s are three Pauli matrices, and $\chi=\mathrm{sgn}(v_1 v_2 v_3)=\pm 1$ captures the chirality or the monopole strength of the Weyl fermions. Due to a ``no-go theorem" of Nielsen-Ninomiya~\cite{Nielsen:1981, Nielsen:1983}, the Weyl points of opposite chirality always come in pairs (except on the surface of a four dimensional topological insulator), and the net monopole charge vanishes. Since the Weyl points of opposite chirality remain separated in the momentum space, the nodal separation vector introduces  a preferred inertial frame. Consequently, a Weyl semimetal always lacks Lorentz invariance~\cite{Goswami:2013} (even if their velocity was equal to the speed of light $c$), despite exhibiting the $z=1$ ($E \sim |\mathbf{k}|$) scaling of energy-momentum relation. The violation of Lorentz invariance and the existence of nontrivial Berry curvature lead to many anomalous transport and optical properties such as large anomalous Hall effect and optical gyrotropy, and the most peculiar one being the negative longitudinal magnetoresistance due to the chiral or Adler-Bell-Jackiw anomaly~\cite{Adler:1969,Bell:1969,Nielsen:1981, Nielsen:1983, Aji:2012, Zyuzin:2012, Volovik, Wan:2011, Xu:2011, Goswami:2013,  Son:2013, Kim:2014, Goswami:2015, GoswamiPixley, Polini:2015, Zubkov:2016}.

In the absence of any gauge or gravitational field coupling, the numbers of right and left handed Weyl fermions is separately conserved. However, in the presence of background gauge fields, such as externally imposed parallel electric and magnetic fields, 
the separate number conservation laws are violated due to subtle quantum mechanical effects~\cite{Adler:1969, Bell:1969}, leaving only the total number to be conserved. This effect is succinctly described by 
$\partial_\mu j^\chi_\mu = -\chi \frac{e^2}{h^2}\mathbf{E}\cdot\mathbf{B}$, and a field-configuration with $\mathbf{E}\cdot\mathbf{B}\neq 0$ can induce charge pumping from one Weyl node ($\chi=1$) to the other ($\chi=-1$) node. An important criterion for the existence of chiral anomaly is the unbounded linear dispersion of the quasiparticles, and in the continuum theory the particles from one Weyl point transfer to the other through the infinite Dirac sea. In a solid state system, in addition to the externally applied electric field there is always a periodic electric field due to  the crystal, and the dispersion relations are bounded. Hence, it will seem impossible to observe any tangible effects of chiral anomaly in any solid state system. But, in the presence of a relaxation mechanism, the scattering rate cuts off the effects of periodic electric field (Bloch oscillations), thus allowing the effects of anomaly to manifest in longitudinal magnetotransport measurements. For weak magnetic fields (when disorder broadening wipes out Landau quantization) semiclassical calculations~\cite{Son:2013, Kim:2014} suggest that $\mathbf{E}\cdot\mathbf{B}$ term can lead to a positive longitudinal magnetoconductance (LMC) while the transverse magnetorsistance remains positive. Similar conclusions are also reached from the calculations in the quantized Landau level basis, particularly in the quantum limit~\cite{GoswamiPixley}. Recently, several experimental groups have found the evidence of chiral anomaly induced positive LMC in Dirac and Weyl materials~\cite{HKim:2013,He:2014, Liang:2015,Xiong:2015,CLZhang:2016,HLi:2016,QLi:2016}.

\begin{figure}
\begin{center}
\includegraphics[scale=0.27]{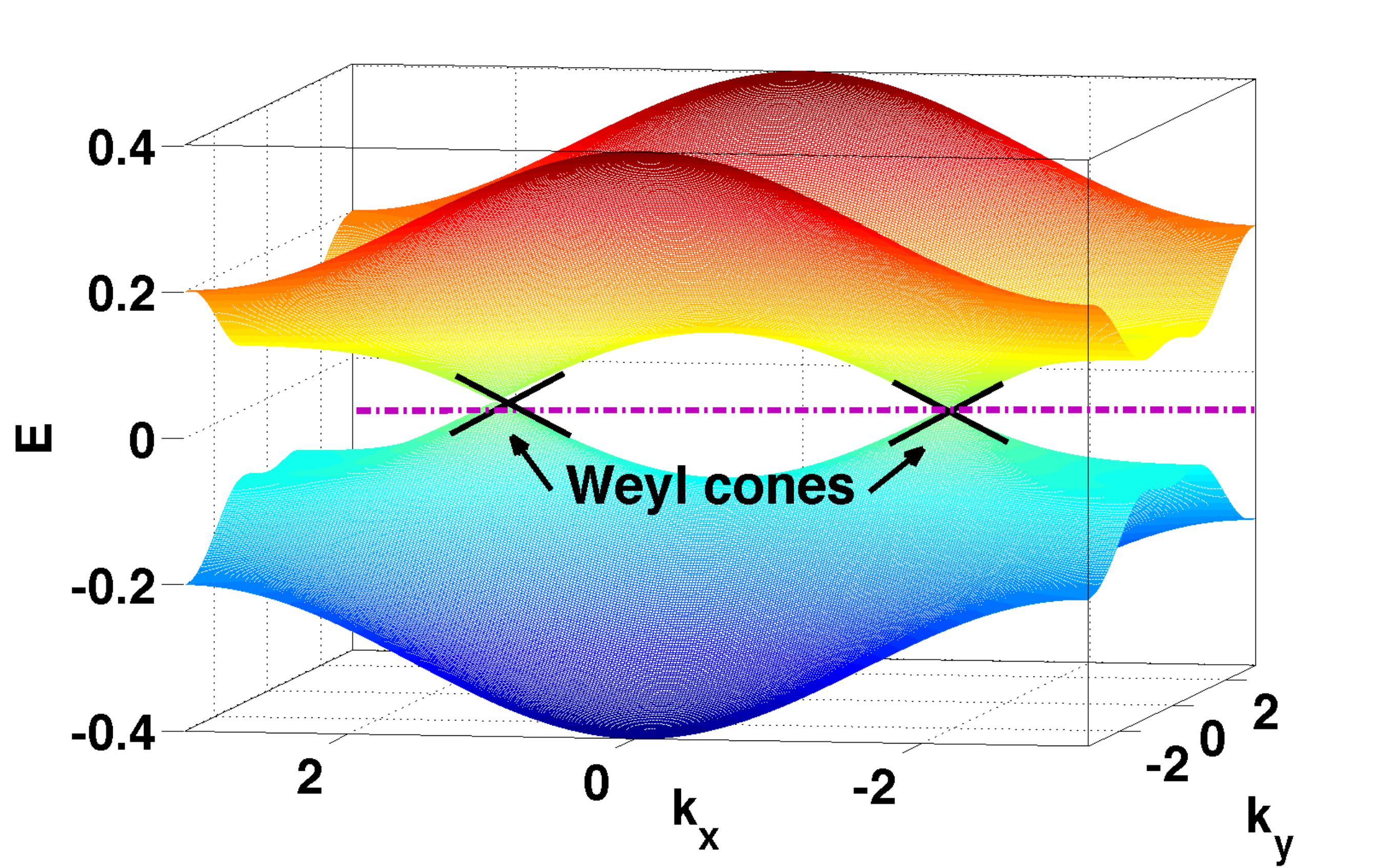}
\includegraphics[scale=0.27]{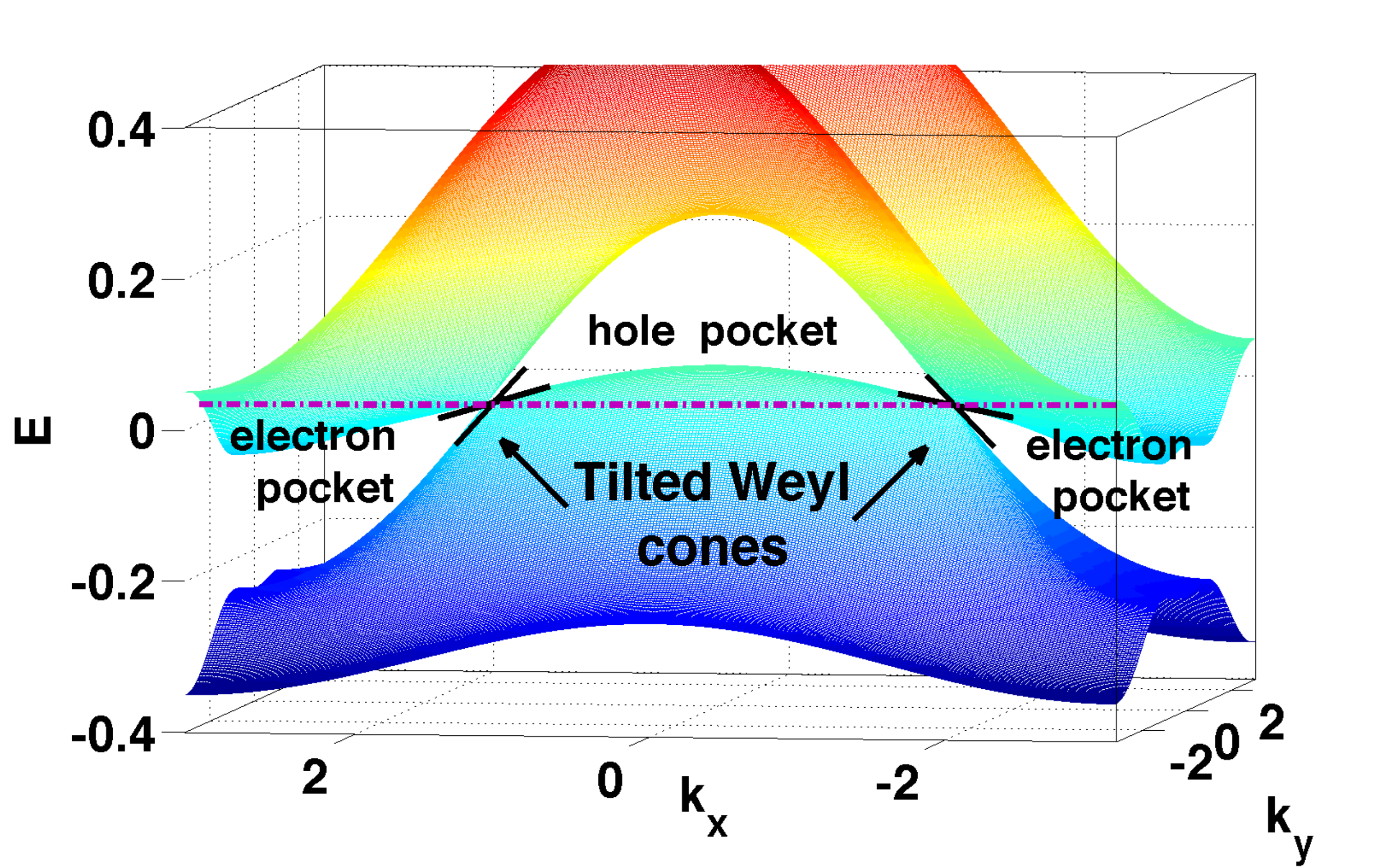}
\caption{Band dispersion for a lattice model of Weyl semimetal governed by Eq~\ref{H_weyl_2_lattice}. \textit{Top panel:} Weyl semimetal of type-I when $\gamma=0$. The Weyl cones are highlighted with black (solid) lines. The chemical potential is placed at $E=0$ indicated by the pink (dashed-dotted) line. Nodal points exist at the intersection of the pink (dashed-dotted) line with the cones. \textit{Bottom panel:} Weyl semimetal of type-II when $\gamma=0.15$ (see Eq.~\ref{H_weyl_2_lattice} for definition of $\gamma$). The Weyl cones around $(\pm k_0,0,0)$ are tilted and the Weyl nodes exist at the boundary of electron and hole pockets. The other parameters used are $m=0.15$, $t=-0.05$, and $2k_0=\pi$. }
\label{Fig_bs}
\end{center}
\end{figure}

For a Weyl node given by Eq.~\ref{Eq_H_k_weyl_1}, the Fermi surface is point like with a conical spectrum along any two dimensions. The spectrum becomes anisotropic when the cone is tilted along a given direction in the momentum space. The Hamiltonian of a linearized tilted Weyl node can be written as \cite{YXu:2015,Soluyanov:2015},
\begin{eqnarray}
H_{\mathbf{k}}^\chi = \chi\hbar v_F (\mathbf{k}-\mathbf{K})\cdot\boldsymbol{\sigma} + \sum\limits_{i\in \{x,y,z\}} (c_ik_i\sigma_0)
\label{Eq_H_k_weyl_2}
\end{eqnarray}
The energy dispersion around the Weyl node $\mathbf{K}$ is now given by $E(\mathbf{k})=\pm\hbar v_F |\mathbf{k}| + T(\mathbf{k})$, with $T(\mathbf{k}) = \sum\limits_i {c_i k_i}$. If the anisotropy is small enough, the Fermi surface of the Weyl node is still point like. However, if along a particular direction $u_{\mathbf{k}}$ in momentum space $T(u_{\mathbf{k}}) > \hbar v_F|\mathbf{k}|$, then a Lifshitz transition leads to a new phase  which has been classified as the Weyl semimetal of type-II~\cite{Soluyanov:2015}. Fig.~\ref{Fig_bs} shows the energy dispersion for a lattice model of Weyl semimetal (type-I and type-II with tilted cones). Unlike the type-I Weyl nodes, the type-II Weyl nodes exist at the boundary of electron and hole 
pockets, and the topological response functions associated with type-II WSM are expected to be different from a type-I WSM. Specifically, it has been suggested \cite{Soluyanov:2015} that on application of an external magnetic field, in a type-II WSM the zeroth chiral Landau level is absent if the magnetic field is applied perpendicular to $u_{\mathbf{k}}$. Therefore, chiral anomaly and the associated LMC are expected to show a strong anisotropy in the direction of the applied magnetic field~\cite{Soluyanov:2015}, i.e. chiral anomaly and LMC is only expected to exist when the magnetic field is directed within a cone around the tilt axis $u_{\mathbf{k}}$. In this work we examine the effects of chiral anomaly on longitudinal magnetotransport in a type II WSM by performing quasi-classical Boltzmann formalism, and show that chiral anomaly induced positive LMC exists along all directions. Experimental signatures of type-II WSM have been reported for Mo$_x$W$_{1-x}$Te$_2$, MoTe$_2$, LaAlGe~\cite{Belopolski:2015, Huang:2016, Xu:2016}, making our study of chiral anomaly and LMC particularly pertinent for upcoming experiments.

\section{Quasi-classical description} The presence of non-trivial Berry curvature is well known to substantially modify electronic properties giving rise to anomalous transport~\cite{Niu:2006, Sundaram:1999}.  In earlier works the topological structure of chiral anomaly was introduced into the Boltzmann formalism~\cite{Son:2013}. A topological $\mathbf{E}\cdot\mathbf{B}$ term appears in the dynamics of quasiparticles which experience a non-vanishing Berry curvature effect (see Eq.~\ref{pdot_eqn}). This term acts as an additional pseudo-force (apart from the standard Lorentz force) and is the source of chiral anomaly.  In the present work we examine anomaly related transport phenomena in a generic Weyl semimetal phase (type-I and type-II) from a quasi-classical Boltzmann formalism~\cite{Sharma:2016, Kim:2014, Lundgren:2014, Son:2013, Spivak:2016}.  The imbalance between two chiral species (left and right handed  Weyl fermions) is equilibrated by backscattering between two Weyl points, and for smooth impurity potentials the backscattering time larger than the forward scattering time, causes positive LMC. We compute the longitudinal conductivity ($\sigma_{uu}$) for a linearized description of WSM, and examine the anisotropy in contributions from the $B-$dependent chiral anomaly term. We then extend this approach to the lattice model of a WSM with a naturalized ultraviolet cutoff.

Since impurities cause typical $\tau\sim 10^{-12}s$, only for intense electric fields $E\sim10^7 V/m$ lattice periodicity effects are important, and they can be safely ignored for small external perturbations and relaxation time scales. Similarly periodic effects of crystal for magnetic field problem are important when lattice constant is comparable to magnetic length, which happens for $B\sim 10^4 -10^5 T$. 
For low magnetic fields ($v \tau \ll l_B$, where $l_B=\sqrt{\hbar/(eB)}$ is the characteristic magnetic length for cyclotron motion, $v$ and $\tau^{-1}$ are respectively the velocity and the impurity scattering rate of the Weyl excitations), when Landau quantization can be neglected, a quasi-classical description of the electron motion remains valid, provided the localization effects due to disorder are not important. Depending on the physical system (as measured by the magnitude of Dingle temperature) this approximation can be valid up to a few Teslas. After incorporating the Berry curvature effects~\cite{Niu:2006, Sundaram:1999}, the equations of motion become
\begin{eqnarray}
\mathbf{\dot{r}}_n &= \mathbf{v}_{n,\mathbf{k}} + \mathbf{\dot{k}}_n\times\mathbf{\Omega}_{n,\mathbf{k}} \label{r_dot_eqn_1} \\
\hbar \mathbf{\dot{k}}_n &= e\mathbf{E} + \frac{e}{c}\mathbf{\dot{r}}_n \times\mathbf{B} \label{p_dot_eqn_1}
\end{eqnarray}
Here $\mathbf{v}_{n,\mathbf{k}}=\nabla_{\mathbf{k}} \epsilon_{n,\mathbf{k}}/\hbar$ is the group velocity ( $\epsilon_{n,\mathbf{k}}$ being the quasiparticle energy dispersion for $n-$th band with $n=\pm$), and $\mathbf{\Omega}_{n,\mathbf{k}}$ represents the Berry curvature for the $n-$th band. By solving these coupled equations (Eq.~\ref{r_dot_eqn_1},~\ref{p_dot_eqn_1}) one obtains 
\begin{align}
&\mathbf{\dot{r}}_n = \mathcal{D}(\mathbf{B},\mathbf{\Omega_{n,\mathbf{k}}}) \left[\mathbf{v}_{n,\mathbf{k}} + \frac{e}{\hbar} (\mathbf{E}\times\mathbf{\Omega}_{n,\mathbf{k}}) + \frac{e}{\hbar c}(\mathbf{v}_{n,\mathbf{k}}\cdot\mathbf{\Omega}_{n,\mathbf{k}})\mathbf{B}\right] \label{rdot_eqn}\\
&\hbar \mathbf{\dot{k}}_n = \mathcal{D}(\mathbf{B},\mathbf{\Omega}_{n,\mathbf{k}}) \left[e\mathbf{E} + \frac{e}{c} (\mathbf{v}_{n,\mathbf{k}}\times \mathbf{B}) + \frac{e^2}{\hbar c}(\mathbf{E}\cdot\mathbf{B})\mathbf{\mathbf{\Omega}_{n,\mathbf{k}}}\right] \label{pdot_eqn}
\end{align}
where $\mathcal{D}(\mathbf{B},\mathbf{\Omega}_{n,\mathbf{k}}) = (1+e (\mathbf{B}\cdot\mathbf{\Omega}_{n,\mathbf{k}})/(\hbar c))^{-1}$. The factor $\mathcal{D}(\mathbf{B},\mathbf{\Omega}_{n,\mathbf{k}})$ also modifies the invariant phase space volume according to $d\mathbf{k} d\mathbf{x}\rightarrow \mathcal{D}(\mathbf{B},\mathbf{\Omega}_{\mathbf{k}}) d\mathbf{k} d\mathbf{x}$~\cite{Duval:2006}. The above equations are generally valid for quasiparticles endowed with Abelian Berry curvature and the $\mathbf{E}\cdot\mathbf{B}$ term in Eq.~\ref{pdot_eqn} captures the effects of chiral anomaly. \emph{This remains valid irrespective of whether the system is a type I or type II WSM, as the Berry curvature is only determined by the coefficients of three Pauli matrices.} 

In the presence of impurity scattering, the semiclassical dynamics of quasiparticles is described by the Boltzmann equation~\cite{Ashcroft}
\begin{eqnarray}
\left( \frac{\partial}{\partial t} + \mathbf{\dot{r}}_n\cdot \nabla_{\mathbf{r}_n} + \mathbf{\dot{k}}_n\cdot\nabla_{\mathbf{k}_n}\right) f_{n,\mathbf{k}} = \mathcal{I}_{\text{coll}}\{f_{n,\mathbf{k}}\}
\label{Eq_boltz_1},
\end{eqnarray}
where $\mathcal{I}_{\text{coll}}\{f_{n,\mathbf{k}}\}$ is the collision integral and $f_{n,\mathbf{k}}$ is the electron distribution function. We will only consider elastic scattering due to the impurities and employ the relaxation time approximation $\mathcal{I}_{\text{coll}}\{f_{n,\mathbf{k}}\} = -\frac{f_{n,\mathbf{k}} - f^0_{n,\mathbf{k}}}{\tau_{n,\mathbf{k}}}=-\frac{g_{n,\mathbf{k}}}{\tau_{n,\mathbf{k}}}$, where $f^0_{n,\mathbf{k}}$ is the equilibrium distribution function, $\tau_{n,\mathbf{k}}$ is the phenomenological relaxation rate, and $g_{n,\mathbf{k}}$ measures the deviation from the equilibrium in the steady state. We will further simplify the calculations by ignoring the explicit momentum dependence of $\tau_n$. In the calculation based on the lattice model of a WSM, we will also assume $\tau_+=\tau_-$. 

 Using Eq.~\ref{rdot_eqn} and~\ref{pdot_eqn}, the Boltzmann equation (Eq.~\ref{Eq_boltz_1}) can be solved for the distribution function $f_{n,\mathbf{k}}$ in order to obtain the conductivity $\sigma_{uu}$ for the configuration when $\mathbf{E}=E\hat{u}$ and $\mathbf{B}=B\hat{u}$, where $\hat{u}$ is an arbitrary direction in real space. The longitudinal conductivity $\sigma_{uu}$ is obtained to be~\cite{Kim:2014,Lundgren:2014,Sharma:2016}
\begin{align}
\sigma_{uu} =\sum\limits_n e^2\int{[d\mathbf{k}]\mathcal{D}_n \(v_u + \frac{eB}{\hbar}\mathbf{\Omega_{n,\mathbf{k}}}\cdot\mathbf{v}_{n,\mathbf{k}}\)^2\tau_n \left(-\frac{\partial f_{eq}}{\partial\epsilon}\right)}
\label{sxx}
\end{align}
where $\mathcal{D}_n \equiv \mathcal{D}(\mathbf{B},\mathbf{\Omega}_{n,\mathbf{k}})$. Comparing the above equation with the regular expression for conductivity~\cite{Ashcroft}, the velocity $v_u$ term is replaced by $v_u + \frac{eB}{\hbar}\mathbf{\Omega_{n,\mathbf{k}}}\cdot\mathbf{v}_{n,\mathbf{k}}$ accounting for chiral anomaly. At zero temperature, $-(\partial f_{eq}/\partial \epsilon) = \delta(\epsilon-\epsilon_{\mathbf{k}})$ in Eq.~\ref{sxx}, which just samples the integrand over the Fermi surface.

\section{Linearized Weyl nodes} We now consider a generic Weyl node with dispersion
\begin{eqnarray}
H^{\chi}_{\mathbf{k}} = \hbar v_F  ({\chi}\mathbf{k}\cdot\mathbf{\sigma} + C k_x),
\label{Eq_H_k_weyl_3}
\end{eqnarray}
where the $C$ is the tilt parameter chosen to be non-zero only along the $k_x$ direction. When $|C| > 1$ $(|C| < 1)$, we have type-II (type-I) Weyl node. The Berry curvature for the Weyl node given above does not depend on the tilt parameter $C$ ($\mathbf{\Omega_{\mathbf{k}}}^\chi = \chi \mathbf{k}/4|\mathbf{k}|^3$).  
\begin{figure}
\begin{center}
\includegraphics[scale=0.215]{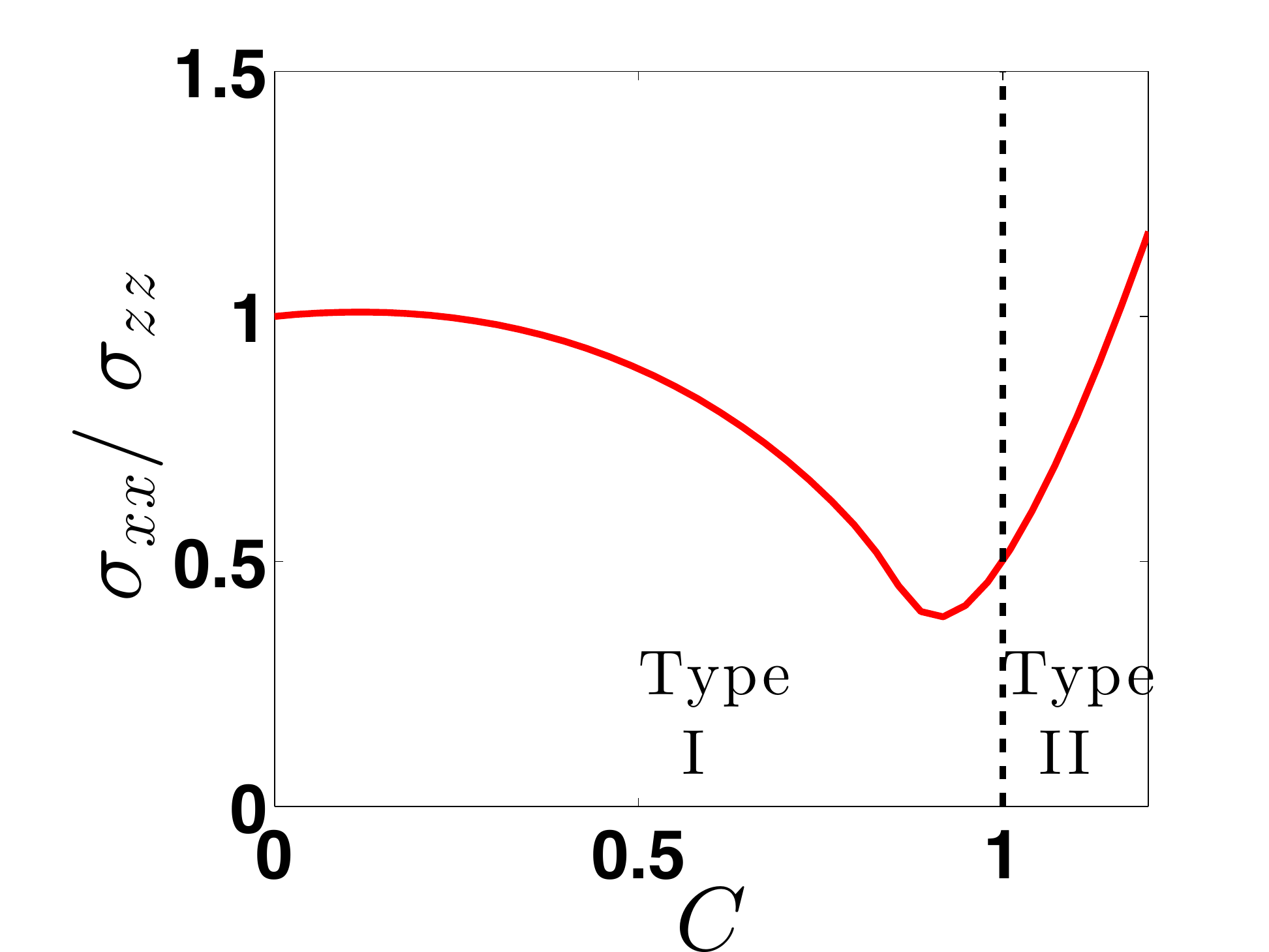}
\includegraphics[scale=0.215]{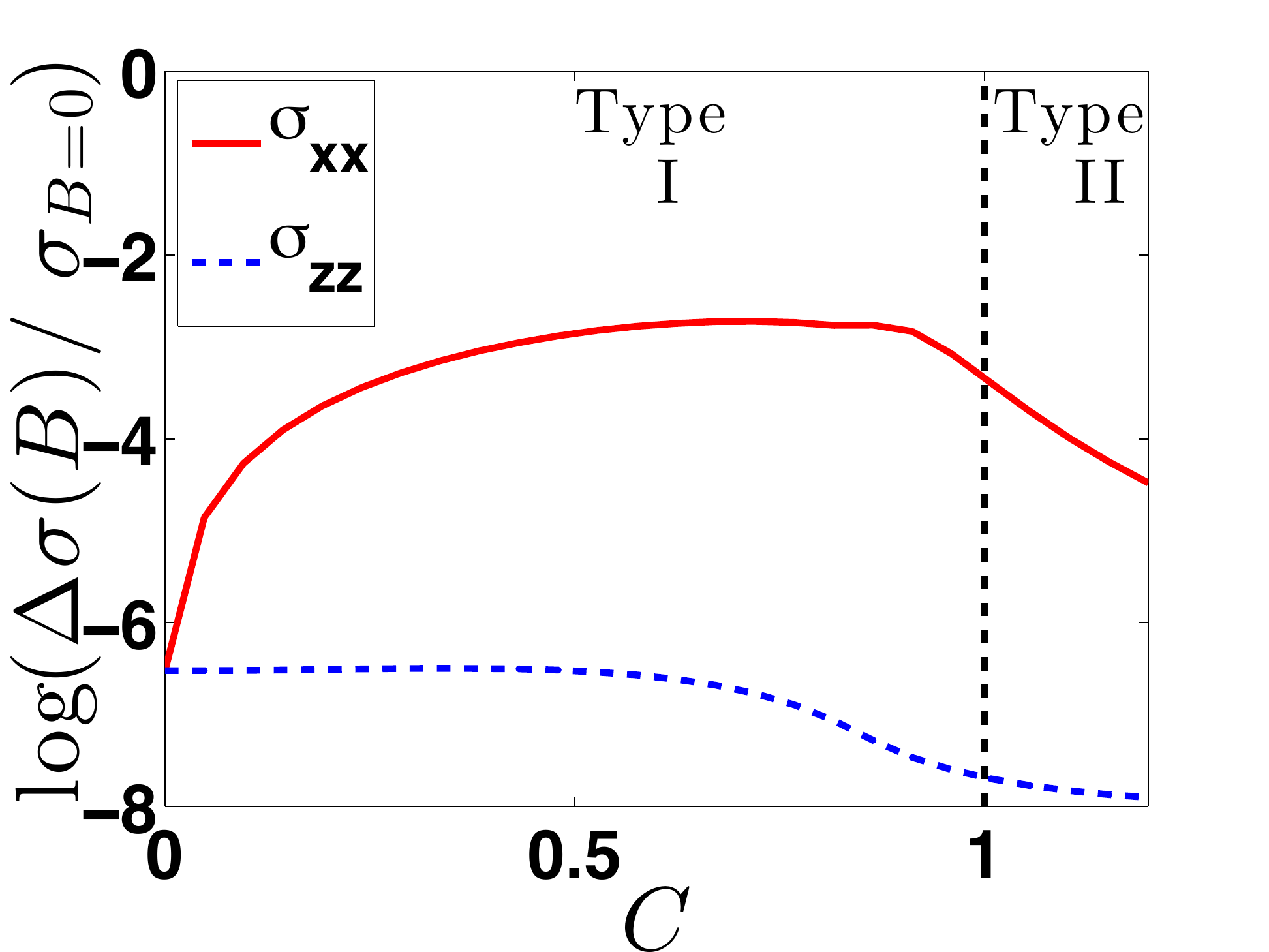}
\caption{\textit{Left:} $\sigma_{xx}/\sigma_{zz}$ numerically computed for a generic linearized Weyl semimetal (Eq.~\ref{Eq_H_k_weyl_3}) with two Weyl nodes, as a function of the tilt parameter $C$, plotted at a magnetic field of $B=0.5T$. For $C=0$, $\sigma_{xx}=\sigma_{zz}$ with no anisotropy. \textit{Right:} $\log(\Delta{\sigma_B}/\sigma_{B=0})$ as a function of tilt parameter $C$. Note that there are no qualitative changes in the behavior of the conductivities at $C=1$, where the Hamiltonian in  Eq.~\ref{Eq_H_k_weyl_3} passes from type I to type II WSM.}
\label{Fig_sigma_xx_by_zz_vs_C}
\end{center}
\end{figure}
\begin{figure}
\begin{center}
\includegraphics[scale=0.21]{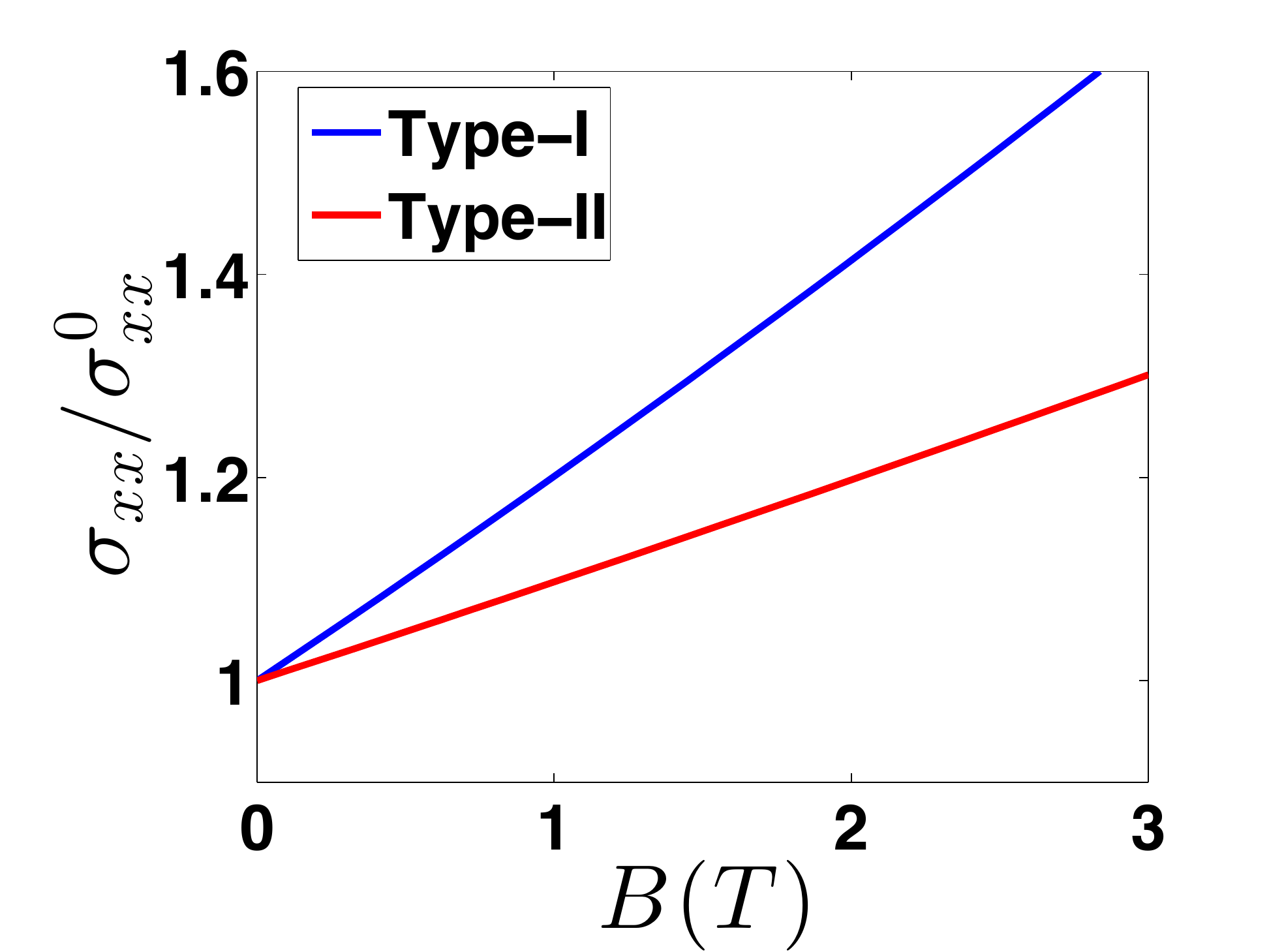}
\includegraphics[scale=0.21]{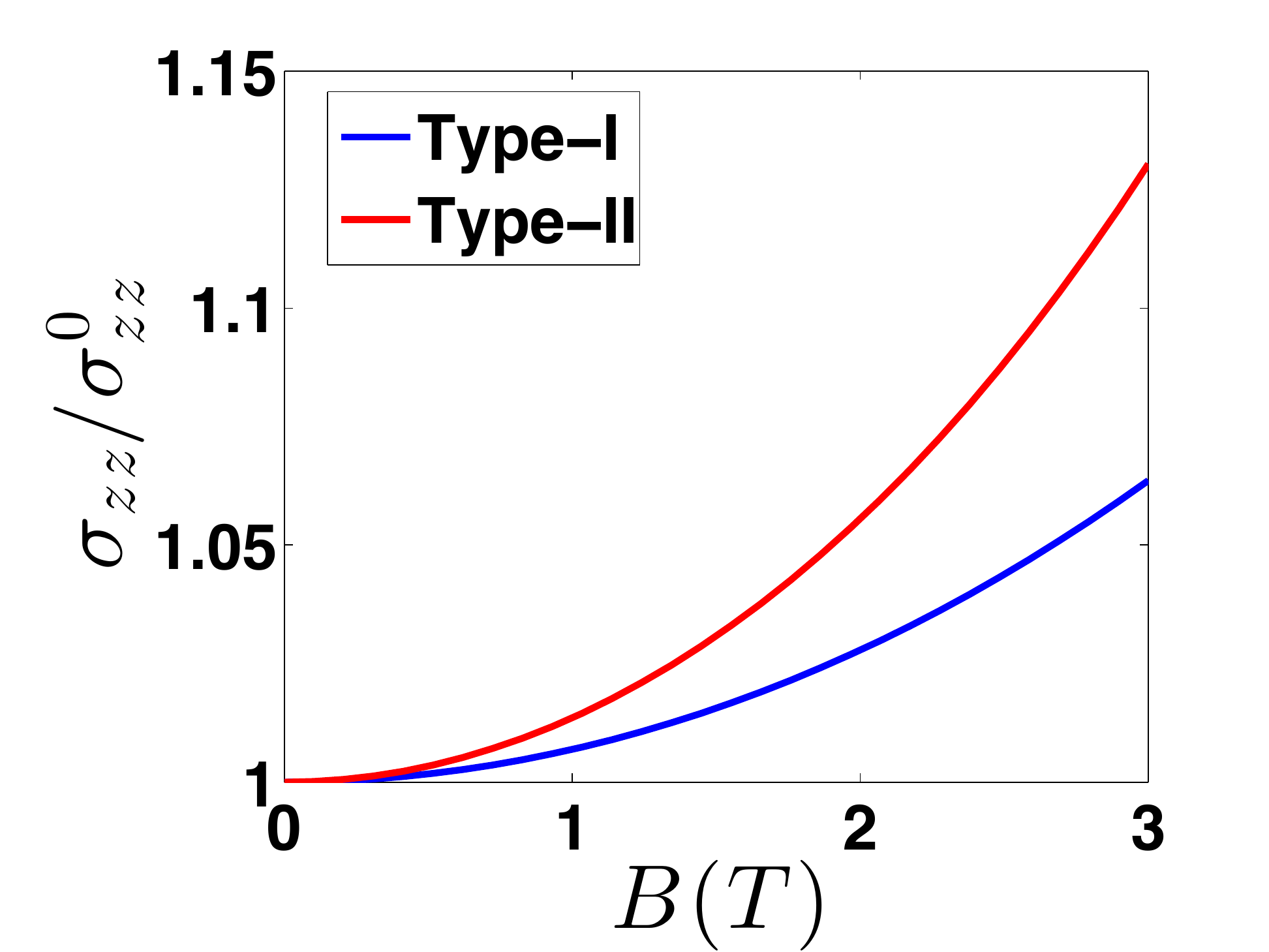}
\caption{Longitudinal magnetoconductivity $\sigma_{uu}(B)$ computed numerically for a linearized Weyl semimetal with two Weyl nodes with $C=\pm 1.2$ (type-II) and $C=\pm 0.7$ (type-I), as a function of the magnetic field of $B$, applied along $x$ and $z$ directions for computation of $\sigma_{xx}$(right) and $\sigma_{zz}$(left) respectively. The conductivity $\sigma^0_{xx/zz}$ is the value at zero magnetic field. }
\label{Fig_sigma_xx_by_zz_vs_B}
\end{center}
\end{figure}
For $C\neq 0$, the Fermi surface at a finite chemical potential is no longer spherical, and is marked with the appearance of Fermi pockets for $|C|>1$. Therefore, analytic evaluation of the conductivity becomes intractable, and we resort to numerical computation of $\sigma_{uu}$. We directly compute the conductivities from Eq.~\ref{sxx}, with an upper ultraviolet cutoff beyond which the linearized description is no longer valid. For numerical computation, the Fermi velocity was chosen to be $v_F=10^6m/s$, and the upper energy cutoff to be $\sim 0.3eV$. Further for our calculations, we consider two Weyl nodes (with chiralities $\chi$ and $-\chi$, and tilt parameters $C$ and $-C$), and add their respective contributions. Figure ~\ref{Fig_sigma_xx_by_zz_vs_C} shows $\sigma_{xx}/\sigma_{zz}$ numerically computed for a generic linearized Weyl semimetal (Eq.~\ref{Eq_H_k_weyl_3}) with two Weyl nodes, as a function of the tilt parameter $C$, plotted at a specific non-zero magnetic field.  
Figure~\ref{Fig_sigma_xx_by_zz_vs_C} also shows $\log(\Delta\sigma(B)/\sigma_{B=0})$, where $\Delta\sigma(B)=(\sigma(B)-\sigma_{B=0})$, for both conductivities along the tilt-direction (i.e. $\sigma_{xx}$) and perpendicular to the tilt direction ($\sigma_{zz}$). We note that there are no qualitative changes in the behavior of the conductivities at $C=1$ where the system changes from type I to type II WSM.


Figure~\ref{Fig_sigma_xx_by_zz_vs_B} shows $\sigma_{xx}$ and $\sigma_{zz}$ as a function of magnetic field $B$. The behavior is quadratic (linear) in $B$ for $\sigma_{zz}$($\sigma_{xx}$) for both type-I and type-II WSMs. This again illustrates the fact that chiral anomaly related positive LMC phenonmena does not differentiate type-I from a type-II WSM, at least within the low field quasiclassical approximation. Further our calculations suggest that the $B-$dependence of LMC is approximately $B-$linear when the applied magnetic field is along the tilt axis, and quadratic in $B$ when the applied magnetic field in perpendicular to the tilt direction.
\begin{figure}
\includegraphics[scale=0.21]{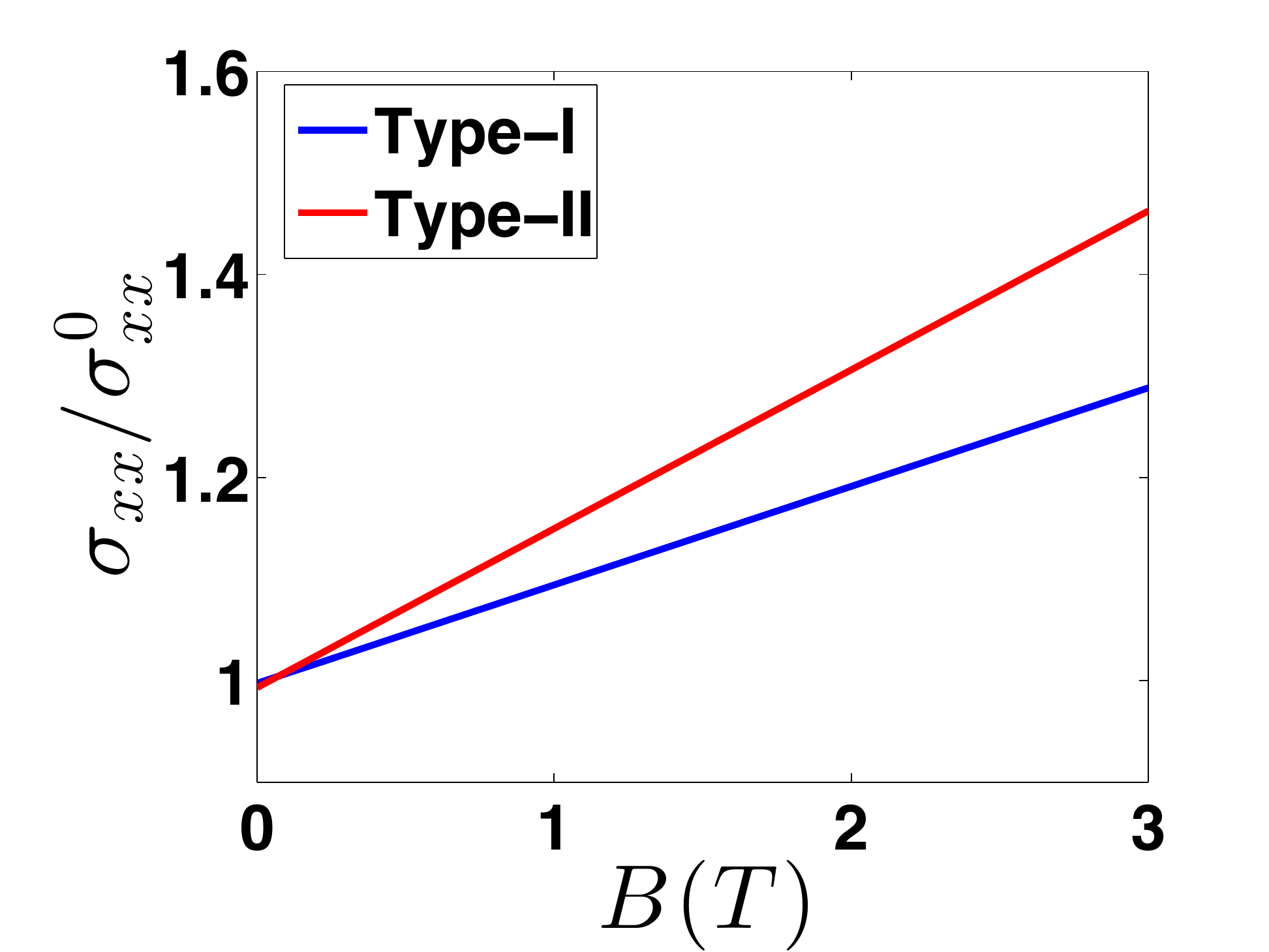}
\includegraphics[scale=0.21]{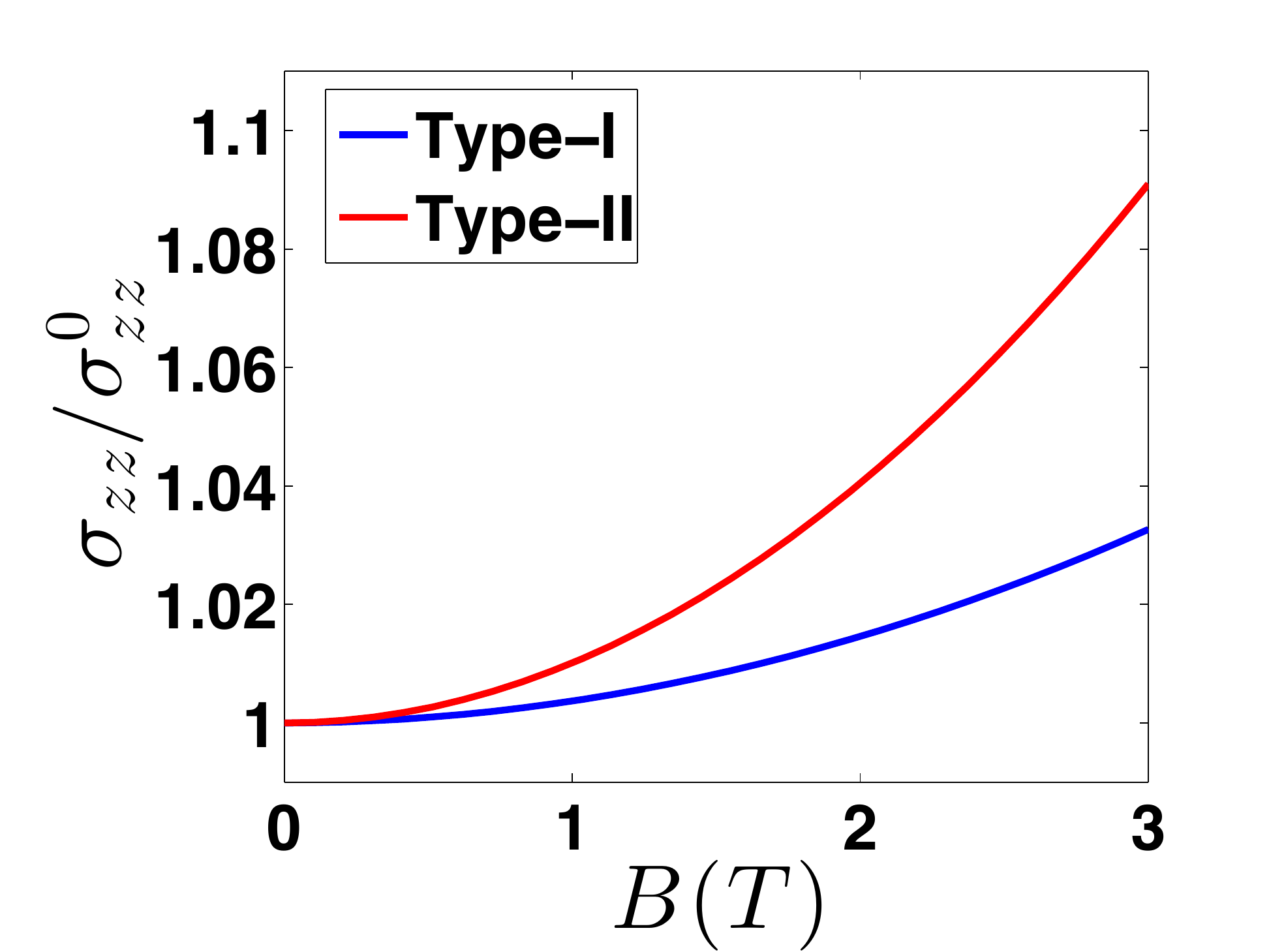}
\includegraphics[scale=0.21]{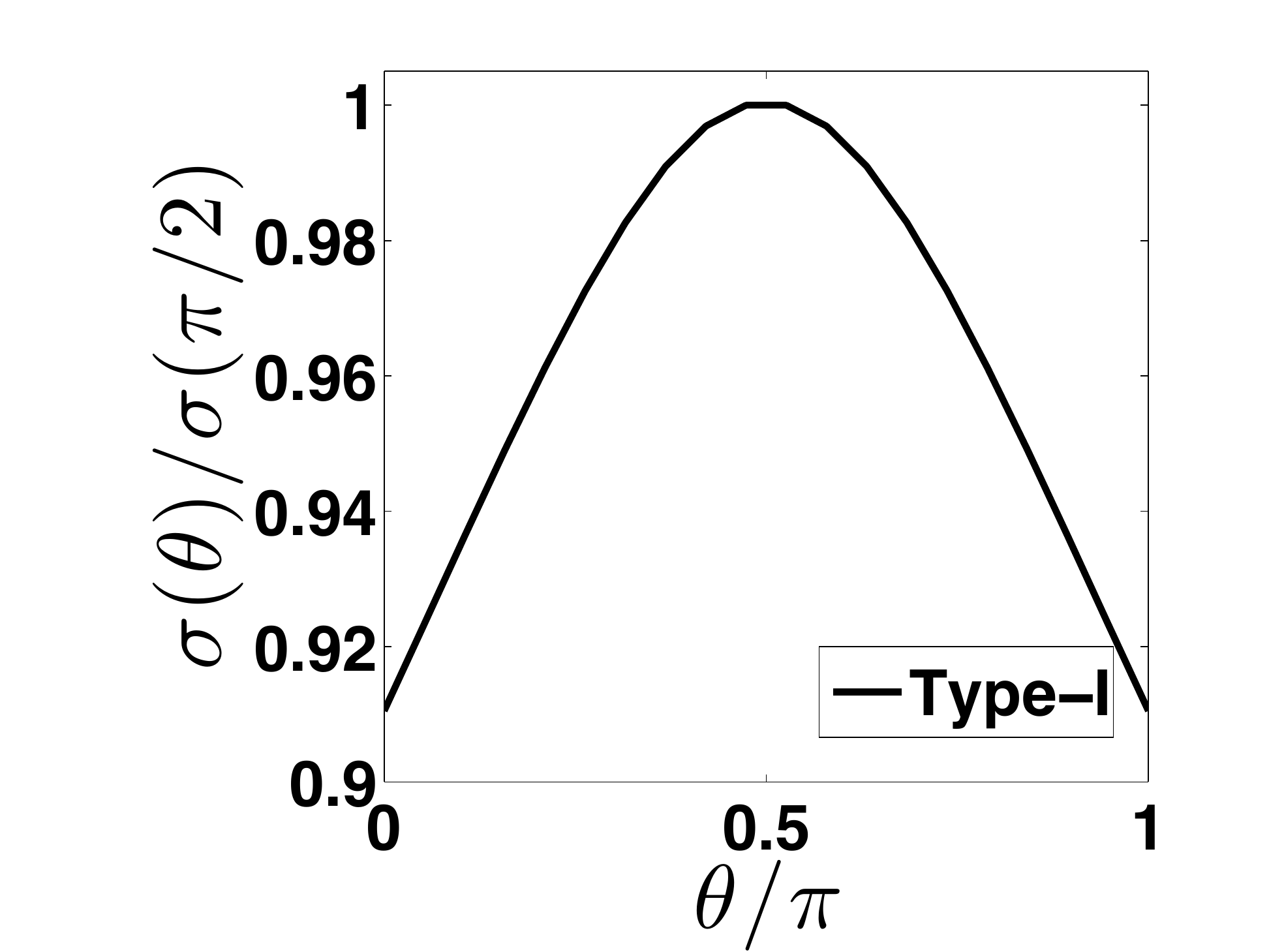}
\includegraphics[scale=0.21]{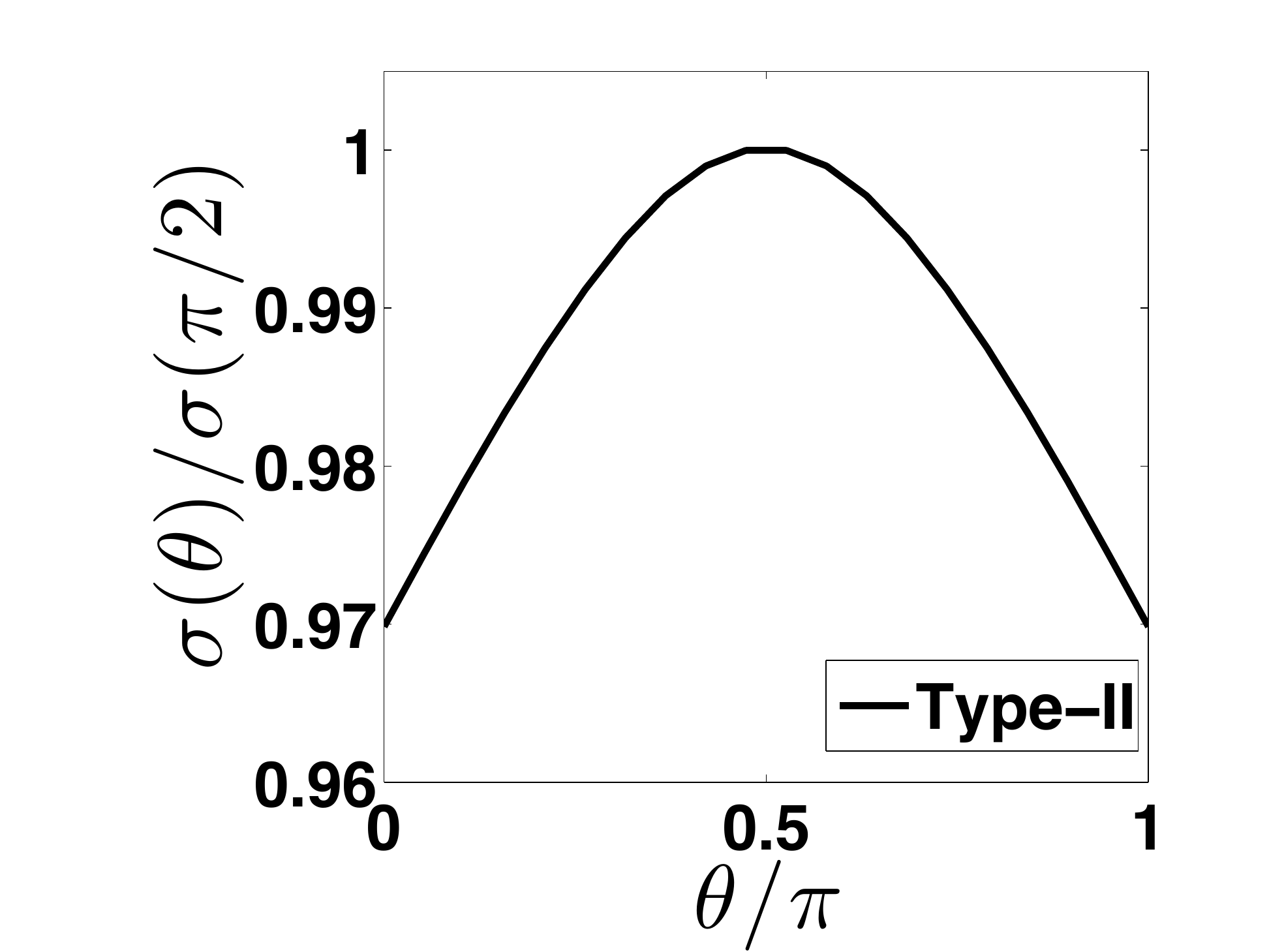}
\caption{\textit{Top panels:} $B-$dependent longitudinal conductivities computed for the lattice model of a Weyl semimetal given by Eq.~\ref{H_weyl_2_lattice}, for $\gamma=0.15$ (type-II) and $\gamma=0.07$ (type-I). For computing $\sigma_{xx}$ (top right panel) and $\sigma_{zz}$ (top left panel) the external $B-$ field is applied along the $x$ and $z$ directions respectively. $\sigma_{xx}$ shows a linear $B-$ dependence, while $\sigma_{zz}$ is quadratic in $B$. Further, $\sigma^0_{xx/zz}$ represent the conductivity at zero magnetic field. \textit{Bottom panels:} $\sigma(\theta)$ (in arbitrary units) as a function of $\theta$ measured from the $z$ axis for a constant magnetic field of magnitude $B=1T$, but rotating in the $xz-$plane from $\theta=0$ to $\theta=\pi$. When $\theta=\pi/2$ the conductivity $\sigma(\pi/2)=\sigma_{xx}$ reaches a maxima. The parameter $t$ was chosen to be $t=-0.05$. Note that these plots essentially drive out the qualitative behavior of LMC and do not make accurate quantitative predictions.}
\label{Fig_sigma_lattice}
\end{figure}
\section{Lattice model of a WSM} It is advantageous to consider a lattice model of Weyl fermions
with the lattice regularization providing a physical ultra-violet smooth cut-off to the low energy spectrum. We now consider a prototype TR-breaking Hamiltonian which produces two Weyl nodes at $\mathbf{K^{\pm}}=(\pm k_0,0,0)$~\cite{Trivedi:2016}
\begin{eqnarray}
H^I(\mathbf{k}) &= ((\cos k_y + \cos k_z - 2)m + 2t(\cos k_x - \cos k_0))\sigma_1 \nonumber \\
&- 2t\sin k_y \sigma_2 - 2t\sin k_z\sigma_3
\end{eqnarray}
The nodes at $\mathbf{K^{\pm}}$ can be tilted in the $k_x$ direction by adding a term as follows
\begin{eqnarray}
H^{II}(\mathbf{k}) = H^I(\mathbf{k}) + \gamma (\cos k_x - \cos k_0)\sigma_0
\label{H_weyl_2_lattice}
\end{eqnarray}
where $\sigma_0$ is an identity matrix. Figure~\ref{Fig_bs} shows the energy dispersion for the lattice model given in Eq.~\ref{H_weyl_2_lattice}. Linearizing near the nodal points, the Hamiltonian $H^{II}(\mathbf{k})$ can be reduced to $H_{lin}(\mathbf{K}^{\pm}\pm\mathbf{k})\approx \mp 2t(\sin k_0 k_x)\sigma_1 - 2t(k_y\sigma_2 + k_z \sigma_3) + \gamma (\mp \sin k_0 k_x)\sigma_0$.
 When $\gamma\neq 0$, the lattice Hamiltonian produces two Weyl nodes which are tilted along the $k_x$ direction and oppositely oriented to each other.
 When $\gamma>|2t|$, the type-II WSM phase emerges, also illustrated in Fig.~\ref{Fig_bs}.

Using Eq.~\ref{sxx} we now compute the $B$-dependent longitudinal conductivities along $\hat{x}$ (parallel) and $\hat{z}$ (perpendicular) directions. Figure~\ref{Fig_sigma_lattice} plots the computed conductivities for the case: $\gamma=0.07$ (type-I) and $\gamma = 0.15$ (type-II). In both cases, $\sigma_{zz}$ has a non-vanishing $B$--dependence (which arises from the chiral anomaly $\mathbf{E}\cdot\mathbf{B}$ term). Thus even if the magnetic field is applied perpendicular to the tilt direction (along $x$ in the present case), one finds a positive LMC. The approximate $B$--dependence along the tilt direction is $B-$ linear. 
 Perpendicular to the tilt direction the $B$--dependence is quadratic. 
 We also plot $\sigma(\theta)$ as a function of $\theta$ measured from the $z$ axis in Figure~\ref{Fig_sigma_lattice}. When $\theta=\pi/2$ the conductivity $\sigma(\pi/2)=\sigma_{xx}$ reaches a maxima, on account of the $B-$linear term.  We therefore conclude that in a type II WSM longitudinal magneto-conductivity is finite at all angles from the tilt direction.

\section{Discussion and Conclusions}
The argument in Ref.~\onlinecite{Soluyanov:2015} is relevant in the strong magnetic field regime when Landau quantization is important. In Ref.~\onlinecite{Soluyanov:2015}, the authors have calculated the Landau level structure within a linearized approximation for $H= C (k_z-eA_z)+ v(\mathbf{k}-e\mathbf{A})\cdot \boldsymbol{\sigma}$, and argue that chiral zeroth Landau level is absent when the magnetic field makes an angle larger than some critical angle determined by the ratio $C/v$. Based on this it has been concluded that chiral anomaly induced LMC should be seen only for a restricted range of angle between the tilt direction and the magnetic field. 
We make a few comments about this calculation: (i) when the angle between the magnetic field and the tilt direction exceeds the threshold, all Landau levels for the above linearized theory disappear (not just the lowest Landau level), which actually capture some pathological properties of the gauged-linearized model. This happens as for a type II system (when $C>v$) as for type I ($C<v$), and the linearized theory does not correctly capture the closed Fermi pockets, from which we are supposed to obtain quantized levels by employing Onsager’s formula. (ii) It is important to retain higher order particle-hole anisotropic terms (which cause tilting) to obtain the correct description of cyclotron orbits or Landau levels. On a qualitative ground consider the situation where particle hole anisotropy is the most dominant term in the Hamiltonian described by $k^2/(2m)$ in an effective mass approximation. In the presence of external magnetic field it produces familiar cyclotron orbits or Landau levels, and spin dependent parts act as small perturbations. Then following the calculations of Ref.~\onlinecite{GoswamiPixley}, one would expect an anomaly induced LMC along all directions for both type I and type II Weyl semimetals. 

In the current work we analyzed WSMs of type-I and type-II using quasi-classical Boltzmann formalism.
Our prediction of a $B-$ linear magneto-conductivity along the direction of tilt in a tilted Weyl semimetal is novel and can be directly tested in experiments. In addition, we prove, using quasiclassical Boltzmann transport theory, that in a type II WSM longitudinal magnetoconductivity is finite at all angles from the tilt direction. In particular, we find that, in contrast to the claims made in Ref.~\onlinecite{Soluyanov:2015}, the LMC is non-zero and quadratic in the applied magnetic field if the latter is applied perpendicular to the tilt direction. In light of a number of recent experiments claiming to have observed type II WSMs, our results on chiral anomaly and longitudinal magneto-conductivity are particularly pertinent for uncovering transport  signatures of type II Weyl semimetals.   

\textit{Acknowledgements.-} GS and ST thank ARO Grant No:
(W911NF-16-1-0182) for support. P. G. was supported by JQI-NSF-PFC and LPS-MPO-CMTC.

\textit{Note Added.-} (i) During the completion of this manuscript we became aware of a recent preprint~\cite{Zyuzin:2016} that also found a $B-$linear magnetoconductivity along the direction of tilt in a tilted Weyl semimetal. (ii) Very recently there has been an experimental observation of Adler-Bell-Jackiw Anomaly in Type-II Weyl semimetal WTe crystals at the quasiclassical regime~\cite{YYLv:2017}, consistent with our theoretical prediction of the existence of B-dependent LMC both perpendicular and parallel to the tilt directions.





\begin{thebibliography}{10}
\bibitem{Peskin} M. E. Peskin, and D. V. Schroeder, \textit{An introduction to quantum field theory}, Westview, (1995).
\bibitem{Murakami1:2007} S. Murakami, New Journal of Physics \textbf{9}, 356 (2007).
\bibitem{Murakami2:2007} S. Murakami, S. Iso, Y. Avishai, M. Onoda, and N. Nagaosa, Phys. Rev. B \textbf{76}, 205304 (2007).
\bibitem{Yang:2011} K. Y. Yang, Y. M. Lu, and Y. Ran, Phys. Rev. B \textbf{84}, 075129 (2011)
\bibitem{Burkov1:2011} A. A. Burkov, M. D. Hook, and L. Balents, Phys. Rev. B \textbf{84}, 235126 (2011).
\bibitem{Burkov:2011} A. A. Burkov and Leon Balents, Phys. Rev. Lett. \textbf{107}, 127205, (2011).
\bibitem{Volovik} G. E. Volovik, Universe in a helium droplet, (Oxford University
Press, 2003).
\bibitem{Wan:2011} X. Wan, A. M. Turner, A. Vishwanath, and S. Y. Savrasov, Phys. Rev. B \textbf{83}, 205101 (2011).
\bibitem{Xu:2011} G. Xu, H. Weng, Z. Wang, X. Dai, and Z. Fang, Phys. Rev. Lett. \textbf{107}, 186806 (2011).
\bibitem{Nielsen:1981} H. B.  Nielsen and M. Ninomiya, Phys. Lett. B \textbf{105} 219 (1981).
\bibitem{Nielsen:1983} H. B. Nielsen and M. Ninomiya, Phys. Lett. B \textbf{130}, 389 (1983).
\bibitem{Goswami:2013} P. Goswami and S. Tewari, Phys. Rev. B \textbf{88}, 245107 (2013).
\bibitem{Adler:1969} S. Adler, Phys. Rev. \textbf{177}, 2426 (1969).
\bibitem{Bell:1969} J. S. Bell and R. A. Jackiw, Nuovo Cimento A \textbf{60}, 47 (1969).
\bibitem{Aji:2012} V. Aji, Phys. Rev. B \textbf{85} 241101 (2012).
\bibitem{Zyuzin:2012} A. A. Zyuzin, S. Wu, and A. A. Burkov, Phys. Rev. B \textbf{85}, 165110 (2012).
\bibitem{Son:2013} D. T. Son and B. Z. Spivak, Phys. Rev. B \textbf{88}, 104412 (2013).
\bibitem{Kim:2014} Ki-Seok Kim, Heon-Jung Kim, and M. Sasaki, Phys. Rev. B \textbf{89}, 195137, (2014).
\bibitem{Goswami:2015} P. Goswami, G. Sharma, S. Tewari, Phys. Rev. B \textbf{92}, 161110 (2015).
\bibitem{GoswamiPixley} P. Goswami, J. H. Pixley, S. Das Sarma, Phys. Rev. B \textbf{92}, 075205 (2015). 
\bibitem{Polini:2015}F. M. D. Pellegrino, M. I. Katsnelson, and M. Polini, Phys. Rev. B \textbf{92}, 201407(R), (2015). 
\bibitem{Zubkov:2016}M. A. Zubkov, Phys. Rev. D, \textbf{93}, 105036 (2016).
\bibitem{HKim:2013}H.-J. Kim, K.-S. Kim, J. F. Wang, M. Sasaki, N. Satoh, A. Ohnishi, M. Kitaura, M. Yang, and L. Li
 , Phys. Rev. Lett. \textbf{111}, 246603 (2013).
\bibitem{He:2014} L. P. He, X. C. Hong, J. K. Dong, J. Pan, Z. Zhang, J. Zhang, and S. Y. Li, Phys. Rev. Lett. \textbf{113}, 246402 (2014).
\bibitem{Liang:2015} T. Liang, Q. Gibson, M. N. Ali, M. Liu, R. J. Cava, N. P. Ong,  Nat Mater \textbf{14}, 280 (2015).
\bibitem{Xiong:2015} J. Xiong, S. K. Kushwaha, T. Liang, J. W. Krizan, M. Hirschberger, W. Wang, R. J. Cava, and N. P. Ong, Science, \textbf{350}, 413 (2015).
\bibitem{CLZhang:2016}C.-L. Zhang, S.-Y. Xu, I. Belopolski, Z. Yuan, Z. Lin, B. Tong, G. Bian, N. Alidoust, C.-C. Lee, S.-M.  Huang, T.-R. Chang, G. Chang, C.-H. Hsu, H.-T. Jeng, M. Neupane, D. S. Sanchez, H. Zheng, J. Wang, H. Lin, C. Zhang, H.-Z. Lu, S.-Q. Shen, T. Neupert, M. Z. Hasan, and S. Jia, Nat. Commun. \textbf{7}, 10735 (2016).
\bibitem{HLi:2016} H. Li, H. He, H.-Z. Lu, H. Zhang, H. Liu, R. Ma, Z. Fan, S.-Q. Shen and J. Wang, Nat. Commun. \textbf{7}, 10301 (2016).
\bibitem{QLi:2016} Q. Li, D. E. Kharzeev, C. Zhang, Y. Huang, I. Pletikosic,
A. V. Fedorov, R. D. Zhong, J. A. Schneeloch, G. D. Gu,
and T. Valla, Nat. Phys. 12, 550 (2016).
\bibitem{YXu:2015}Y. Xu, F. Zhang, and, C. Zhang, Phys. Rev. Lett. \textbf{115}, 265304 (2015).
\bibitem{Soluyanov:2015} A. A. Soluyanov, D. Gresch, Z. Wang, Q. Wu, M. Troyer,
X. Dai, and B. A. Bernevig, Nature \textbf{527}, 495 (2015).
\bibitem{Belopolski:2015} I. Belopolski, S.-Y. Xu, Y. Ishida, X. Pan, P. Yu, D. S.
Sanchez, M. Neupane, N. Alidoust, G. Chang, T.-R.
Chang, Y. Wu, G. Bian, H. Zheng, S.-M. Huang, C.-C. Lee,
D. Mou, L. Huang, Y. Song, B. Wang, G. Wang, Y.-W.
Yeh, N. Yao, J. Rault, P. Lefevre, F. Bertran, H.-T. Jeng,
T. Kondo, A. Kaminski, H. Lin, Z. Liu, F. Song, S. Shin,
and M. Z. Hasan, arXiv:1512.09099, (2015), .
\bibitem{Huang:2016} L. Huang, T. M. McCormick, M. Ochi, Z. Zhao,
M. to Suzuki, R. Arita, Y. Wu, D. Mou, H. Cao, J. Yan,
N. Trivedi, and A. Kaminski, arXiv:1603.06482, (2016).
\bibitem{Xu:2016} S.-Y. Xu, N. Alidoust, G. Chang, H. Lu, B. Singh, I. Belopolski,
D. Sanchez, X. Zhang, G. Bian, H. Zheng, M.-
A. Husanu, Y. Bian, S.-M. Huang, C.-H. Hsu, T.-R.
Chang, H.-T. Jeng, A. Bansil, V. N. Strocov, H. Lin,
S. Jia, and M. Z. Hasan, arXiv:1603.07318 (2016).
\bibitem{Niu:2006} Di Xiao, Yugui Yao, Zhong Fang, and Qian Niu, Phys. Rev. Lett, \textbf{97}, 026603 (2006).
\bibitem{Sundaram:1999} G. Sundaram, and Qian Niu, Physical Review B \textbf{59}, 14915 (1999).
\bibitem{Lundgren:2014} R. Lundgren, P. Laurell, and G. A. Fiete, Phys. Rev. B \textbf{90} 165115 (2014).
\bibitem{Sharma:2016} G. Sharma, P. Goswami, and S. Tewari, Phys. Rev. B \textbf{93}, 035116 (2016).
\bibitem{Spivak:2016} B. Z. Spivak, and A. V. Andreev, Phys. Rev. B \textbf{93}, 085107 (2016).
\bibitem{Duval:2006} C. Duval, Z. Horvth, P. A. Horvthy, L. Martina, and P. C. Stichel, Mod. Phys. Lett. B, \textbf{20}, 373 (2006).
\bibitem{Ashcroft} John. M. Ziman, \textit{Electrons and phonons: the theory of transport phenomena in solids}. Oxford, UK: Clarendon Press, (2001).
\bibitem{Trivedi:2016} T. M. McCormick, I. Kimchi, N. Trivedi, arXiv:1604.03096 (2016).
\bibitem{Zyuzin:2016} 	V. A. Zyuzin, arXiv:1608.01286 (2016).
\bibitem{YYLv:2017} Y-Y. Lv, X. Li, B-B. Zhang, W. Y. Deng, S-H. Yao, Y.B. Chen, J. Zhou, S-T. Zhang, M.-H. Lu, L. Zhang, M.Tian, L. Sheng, and Y-F. Chen, Phys. Rev. Lett. \textbf{118}, 096603 (2017).
%
\end{thebibliography}
\end{document}